\documentclass[12pt]{article}
\usepackage{epsf}
\author{ A.\,Samsonov\thanks{e-mail:sams@heron.itep.ru}\\\\
\it {\small{Institute of Theoretical and Experimental Physics,}} \\
\it {\small{Bol'shaya Cheremushkinskaya, 25, Moscow, 117259, Russia}}}
\bf
\title{Magnetic moment of the $\rho$-meson in QCD sum rules: \\
$\alpha_s$-corrections}

\begin{document}
\date{}
\maketitle
\newcommand{\qq}{\langle\overline{q}q\rangle^2}

\def\la{\mathrel{\mathpalette\fun <}}
\def\ga{\mathrel{\mathpalette\fun >}}
\def\fun#1#2{\lower3.6pt\vbox{\baselineskip0pt\lineskip.9pt
\ialign{$\mathsurround=0pt#1\hfil##\hfil$\crcr#2\crcr\sim\crcr}}}

$$\centerline{\hbox{Abstract}}$$
\rm
\indent The magnetic moment $\mu$ of the $\rho$-meson 
is calculated in the framework of 
QCD sum rules in external fields. $\alpha_s$-corrections are  
taken into account.  
Bare loop calculations (parton model) give: $\mu_{part}=2.0$.
The nonperturbative effects (operators of dimension 6) result in negative
contribution, whereas perturbative corrections -- positive one. 
The final answer is $\mu=1.8\pm 0.3$.
\\
\newpage
%$\,$\\
%\large{
\section{Introduction}
$\,$
\indent QCD sum rule technique is an effective model independent tool in the
study of various characteristics of hadron states. This technique can be
applied in particular to the investigation of the 
static properties of vector mesons. \\
\indent Namely, in the study of photoproduction, 
scattering, etc. of the vector mesons the vector dominance 
hypothesis is used. In the framework of 
this hypothesis it is supposed that the interaction of real
or virtual photon with hadrons proceeds in such a way that the photon 
first transforms into vector mesons $\rho,\,\omega,\,\phi,$ which
then undergo interaction with hadrons. In the consistent lagrangian
formulation of  vector dominance model it is assumed \cite{kroll} 
(see \cite{iof-kh-lip} for review) that $\rho$-mesons are 
Yang-Mills vector bosons. In  this hypothesis
the $\rho$-meson magnetic moment is equal to 2 
(in units $eh/(2m_\rho c)$), at least if strong interaction is 
neglected. \\
\indent In paper \cite{iof-sm-meson} the $\rho$-meson formfactors were found 
 at intermediate momentum transfer by QCD sum rules. By extrapolation
of the $\rho$-meson magnetic formfactor to the point $Q^2=0$ (outside
the applicability domain of the technique) it was found that the $\rho$-meson 
magnetic moment $\mu$ is close to 2. However, this result can not be 
considered as final, and direct calculation of the magnetic moment
is the topical problem. \\
\indent In our previous paper \cite{sams-mrho} the $\rho$-meson magnetic 
moment was calculated in the framework of QCD sum rules in external fields.
This technique was developed in \cite{iof-sm-magmom},\cite{bal-yung}. 
The operators of dimension 6 were taken
into account. As for $\alpha_s$-corrections, they were neglected.
However, since the discussed energy scale is approximately $1\,GeV^2$, 
$\alpha_s$-corrections may give appreciable contribution. \\
\indent This paper is devoted to the calculation 
of the $\alpha_s$-corrections to the $\rho$-meson magnetic moment.
It is organized as follows. 
First of all, we are going to remind the reader the basic points 
of the sum rule construction (section 2). In
the section 3 we discuss in detail the perturbative contribution 
to the sum rule. In the section 4 we present the nonperturbative 
contribution. The results can be found in the section 5. The values 
obtained in some other approaches
(Dyson-Schwinger equation based models \cite{hawes},
\cite{hecht}, relativistic quantum mechanics \cite{rel-qm}, light cone 
QCD sum rules \cite{aliev-kanik}) are also presented. \\
\indent In our paper $u$- and $d$-quarks are considered as massless.
%ttttttttttttt
\section{Phenomenological part of the sum rule}
$\;$
\indent Let us consider the correlator of two vector currents in the external
electromagnetic field:
%1
$$\Pi_{\mu\nu}(p)=i\int d^4x\,e^{ipx}\langle T(j_\mu(x)j^+_\nu(0)\rangle_F\,.
\eqno(1) $$
Here subscript $F$ denotes the presence of the external electromagnetic
field with strength $F_{\rho\lambda}$ and $j_\mu$ is the vector current 
with $\rho$-meson quantum numbers:
$j_\mu=\overline{u}\gamma_\mu d$. Its matrix element is 
%2
$$\langle\rho^+\vert j_\mu \vert 0\rangle =(m_\rho^2/g_\rho)e_\mu\,,
\eqno(2)$$
where $m_\rho$ is the $\rho$-meson mass, $g_\rho$ is the $\rho$-$\gamma$ 
coupling constant, $g_\rho^2/(4\pi)=1.27$, and $e_\mu$ is the $\rho$-meson 
polarization vector. \\
\indent In the limit of weak external field we consider only linear in 
$F_{\rho\lambda}$ terms in the correlator $\Pi_{\mu\nu}$ (1):
%3
$$\Pi_{\mu\nu}=\Pi^0_{\mu\nu}+i\sqrt{4\pi\alpha}
\Pi_{\mu\nu\chi\sigma}F_{\chi\sigma}\,.\eqno(3)$$
\indent In order to find the $\rho$-meson magnetic moment we construct 
sum rule for the invariant  function 
$\Pi(p^2)$ at certain kinematical structure of 
$\Pi_{\mu\nu\chi\sigma}$ (3). \\
\indent The description of the 
kinematical structure choose and sum rule construction  
can be found in  \cite{sams-mrho}.
For the sake of consistency we present it here.\\
\indent The electromagnetic vertex of the $\rho$-meson has the following
general form \cite{iof-sm-meson}:
%4
$$\displaylines{\langle\rho(p+q,e^{r^\prime})|j^{el}_\chi|\rho(p,e^r)\rangle =
-e^{r^\prime}_\sigma e^r_\rho \bigg(\Big((2p+q)_\chi g_{\rho\sigma}-
(p+q)_\rho g_{\chi\sigma}-p_\sigma g_{\rho\chi}\Big)F_1(-q^2)+\hfill}$$
$$\displaylines{\hfill+(g_{\chi\rho}q_\sigma-g_{\chi\sigma}q_\rho)
F_2(-q^2)+{1\over{m_\rho^2}}
(p+q)_\rho p_\sigma(2p+q)_\chi F_3(-q^2)\bigg)\,.~~~~(4)}$$
In this expression 
$j^{el}_\chi=e_u\overline u\gamma_\chi u+e_d\overline d\gamma_\chi d$
is the electromagnetic current, $e_u\,,e_d$ are $u$- and $d$-quark charges 
and $F_1\,,F_2\,,F_3$ are electric, magnetic and quadrupole formfactors
correspondingly. The $\rho$-meson magnetic moment $\mu$ is related to 
magnetic formfactor at $q^2=0$:
%5
$$\mu=1+F_2(0)\,.\eqno(5)$$
\indent Using (2) and (4), we obtain for the 
$\langle 0|j_\mu|\rho\rangle\langle
\rho|j_\chi|\rho\rangle\langle\rho|j_\nu|0\rangle\epsilon_\chi$
transition:
%6
$$\displaylines{
-i\sum\limits_{r,r^\prime}\langle 0|j_\mu|\rho^{r^\prime}\rangle
\langle\rho^{r^\prime}|j^{el}_\chi|\rho^r\rangle\langle\rho^r|
j_\nu|0\rangle\epsilon_\chi=\hfill}$$
$$\displaylines{\hfill=i{m_\rho^4\over{g_\rho^2}}\sum_{r,r^\prime}
 e^{r^\prime}_\mu e^{r^\prime}_\sigma e^r_\rho e^r_\nu \epsilon_\chi
\bigg(\Big((2p+q)_\chi g_{\rho\sigma}-
(p+q)_\rho g_{\chi\sigma}-p_\sigma g_{\rho\chi}\Big)F_1(-q^2)+
~~~~~~(6)}$$
$$\displaylines{\hfill+(g_{\chi\rho}q_\sigma-g_{\chi\sigma}q_\rho)
F_2(-q^2)+{1\over{m_\rho^2}}
(p+q)_\rho p_\sigma(2p+q)_\chi F_3(-q^2)\bigg)\,.~~~~}$$
Here  $\epsilon_\chi-$photon polarization, $r,\,r^\prime$ are 
the $\rho$-meson polarization indices. Let us consider in this 
expression linear in $q_\sigma$ terms. We sum over 
$\rho$-meson polarizations, retain the antisymmetric over $\chi,\sigma$ part,
introduce the electromagnetic field $F_{\chi\sigma}=
i(\epsilon_\chi q_\sigma-\epsilon_\sigma q_\chi)$ and obtain for (6):
$$\displaylines{-{m_\rho^4\over{2g_\rho^2}}F_{\chi\sigma}
\bigg((F_2+{1\over 2}F_1){1\over{p^2}}\Big(p_\nu
(p_\chi g_{\mu\sigma}-p_\sigma g_{\mu\chi})-p_\mu
(p_\chi g_{\nu\sigma}-p_\sigma g_{\nu\chi})\Big)+\hfill}$$
$$\displaylines{\hfill+{1\over 2}F_1{1\over{p^2}}\Big(p_\nu
(p_\chi g_{\mu\sigma}-p_\sigma g_{\mu\chi})+p_\mu
(p_\chi g_{\nu\sigma}-p_\sigma g_{\nu\chi})\Big)+(F_2+F_1)
(g_{\mu\chi}g_{\nu\sigma}-g_{\mu\sigma}g_{\nu\chi})\bigg)\,.}$$
Formfactor $F_3$ does not give linear in $q_\sigma$ contribution. \\
\indent The $\rho$-meson magnetic moment $\mu$ is involved in the invariant 
functions at the kinematical structures
$p_\nu(p_\chi g_{\mu\sigma}-p_\sigma g_{\mu\chi})-
p_\mu(p_\chi g_{\nu\sigma}-p_\sigma g_{\nu\chi})$ and 
$g_{\mu\chi}g_{\nu\sigma}-g_{\mu\sigma}g_{\nu\chi}$.
We choose the structure
%7
$$p_\nu(p_\chi g_{\mu\sigma}-p_\sigma g_{\mu\chi})-
p_\mu(p_\chi g_{\nu\sigma}-p_\sigma g_{\nu\chi})\,.\eqno(7)$$
In comparison with another one, (7) contains two additional powers 
of momentum in the numerator, which result in better convergence of the 
operator expansion series. \\
%\indent It is worth to note here that,
%as follows from the vector current 
%conservation, the antisymmetric over field indices $\chi,
%\sigma$ structure in $\Pi_{\mu\nu\chi\sigma}$ (3) is 
%antisymmetric over $\rho$-meson indices $\mu,\nu$ too. \\  
\indent Using equalities (5) and $F_1(0)=1$ one can see that
$F_2(0)+1/2F_1(0)=\mu-1/2$. \\
\indent Thus, one should calculate the invariant function $\Pi(p^2)$ 
at the structure (7) in $\Pi_{\mu\nu\chi\sigma}$. \\
\indent In order to obtain the phenomenological part of the sum rule, 
one should
saturate the dispersion relation for $\Pi(p^2)$ by the contribution of
physical states. We use the simplest model of physical spectrum, 
which contains the lowest resonance and continuum.
Phenomenological representation of $\Pi$ has the form:
$$\Pi(p^2)=\int ds{\rho(s)\over{(s-p^2)^2}}+...
\,,$$
$$\rho(s)=-{m_\rho^4\over{2g_\rho^2}}{1\over s}\Big(\mu-{1\over 2}\Big)
\delta(s-m_\rho^2)+
f(s)\theta(s-s_\rho)\,.$$
Here dots mean the contributions of non-diagonal 
transitions (for example, \\ 
$\langle 0|j_\mu|\rho^\star\rangle\langle
\rho^\star|j_\chi \epsilon_\chi|\rho\rangle\langle\rho
|j_\nu|0\rangle$, where $\rho^\star$ is the excited state with 
the same quantum numbers as $\rho$), 
function $f(s)$ represents continuum contribution 
and $s_\rho$ is the continuum threshold for the $\rho$-meson.\\
\indent Retaining only the terms, 
which do not vanish after Borel transformation, we obtain:
%8
$$\Pi(p^2)=-{m_\rho^2\over{2g_\rho^2}}
{{\mu-{1\over 2}}\over{(m_\rho^2-p^2)^2}}+
{C\over{m_\rho^2-p^2}}+
\int\limits_{s_\rho}^\infty ds{f(s)\over{(s-p^2)^2}}\,,\eqno(8)$$
where $C$ appears due to non-diagonal transitions.
%ttttttttttt
\section{Perturbative contribution to the sum rule}
$\,$ \indent Now we need to calculate $\Pi(p^2)$, basing on the
operator product expansion in QCD. In this section we consider perturbative
terms. \\
\begin{figure}[h]
\epsfxsize=3.0cm
\epsfbox{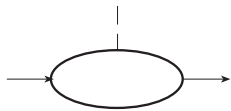}
\caption{\small{Loop diagram in the external electromagnetic field. 
Thick solid lines
denote quarks, dashed line denotes the external field.}}
\end{figure}
\begin{figure}[h]
\epsfxsize=11.0cm
\epsfbox{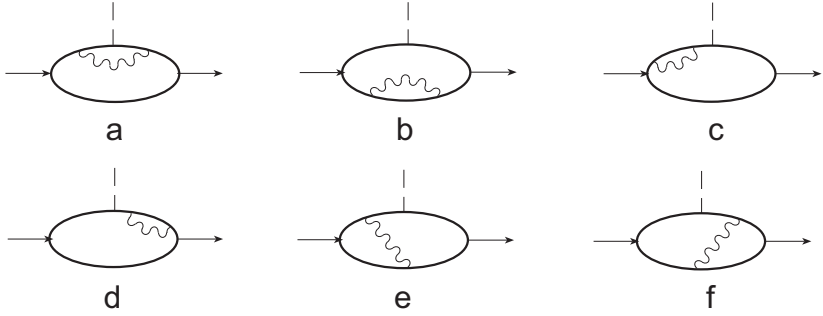}
\caption{\small{Perturbative corrections to the bare loop. 
Wave lines denote gluons.}}
\end{figure}
\indent The loop diagram (fig.1) corresponds to the operator of the lowest
dimension 2. \\Its contribution to $\Pi(p^2)$ is equal to
$$-{3\over{16\pi^2}}\int\limits_0^\infty{ds\over{(s-p^2)^2}}\,.$$
\indent The expressions for the quark propagator
in the external electromagnetic field and in the  
external electromagnetic and soft gluon fields can be found in 
\cite{iof-sm-magmom}. \\
\indent In order to improve the free-quark approximation, we calculate the 
$\alpha_s$ perturbation corrections corresponding to the diagrams in fig.2. 
We consider the integrals with respect to internal momenta 
in $D=4-2\epsilon$ dimensions. 
For the calculation of the diagrams in fig.2e and fig.2f the following 
formula from the paper \cite{chetyrkin} is used:
$$\displaylines{\epsilon\int{d^Dk\,d^Dq\over{(2\pi)^{2D}}}
{1\over{k^2(p-k)^2q^2(q-k)^2(p-q)^2}}=\hfill}$$
$$\displaylines{\hfill=\int{d^Dk\,d^Dq\over{(2\pi)^{2D}}}
\Bigg({1\over{k^2(p-k)^4q^2(k-q)^2}}-
{1\over{k^2(p-k)^4q^2(p-q)^2}}\Bigg)\,.}$$
In the limit 
$\epsilon\to 0$ these diagrams give the following contributions to the 
polarization operator:
$$\displaylines{
A_a={\alpha_s\over{2^6\pi^3p^2}}\Big({4\over\epsilon}+22-8\gamma_E\Big)
~~~~~~~(\hbox{diagram in fig.2a)}\,,\hfill}$$
$$\displaylines{
A_b={\alpha_s\over{2^6\pi^3p^2}}\Big(-{4\over\epsilon}-10+8\gamma_E\Big)
~~~~~~~(\hbox{diagram in fig.2b)}\,,\hfill}$$
$$\displaylines{
A_c=A_d={\alpha_s\over{2^6\pi^3p^2}}\Big(-{4\over\epsilon}-14+8\gamma_E\Big)
~~~~~~~(\hbox{diagrams in fig.2c,d)}\,,\hfill}$$
$$\displaylines{
A_e=A_f={\alpha_s\over{2^6\pi^3p^2}}\Big({4\over\epsilon}+{44\over 3}
-8\gamma_E\Big)~~~~~~~(\hbox{diagrams in fig.2e,f)}\,.\hfill}$$
Here $\gamma_E$ is the Euler constant. \\
\indent In the sum 
$$A_a+A_b+A_c+A_d+A_e+A_f={5\alpha_s\over{24\pi^3p^2}}$$
the divergence $1/\epsilon$ cancels out.
It is not by chance, since the diagrams in fig.2 correspond 
to the physically measurable process, and their
total contribution can not include divergent term. \\
\indent After adding the 
$\alpha_s$-correction, we  obtain for the perturbative (and bare loop) 
contribution to the polarization operator:
%9
$$-{3\over{16\pi^2}}\Big(1+{10\over 9}{\alpha_s\over\pi}\Big)
\int\limits_0^\infty{ds\over{(s-p^2)^2}}\,.\eqno(9)$$
\indent According to the quark-hadron duality, 
the continuum contribution in the
interval of $P^2=-p^2$ from $s_\rho$ to infinity is determined by 
term (9) in this interval. 
Therefore, function $f$ in (8) is: 
$f=-3/(16\pi^2)(1+10\alpha_s/(9\pi))$.
%ttttttttttt
\section{Contribution of dimension 6 operators}
$\,$ \indent The contribution of the operators of dimension 6 was found in 
\cite{sams-mrho}. Here we remind the corresponding formulas. \\ 
\indent The diagrams in fig.1,2 
correspond to the operator of the lowest dimension
$F_{\rho\lambda}$. The operators of dimension 4 are absent. As was shown in 
\cite{iof-sm-magmom}, operator $\overline{q}(D_\mu \gamma_\nu -
D_\nu \gamma_\mu)q$ has opposite with respect to the electromagnetic field
C-parity and can not be induced by them, while operator
$\epsilon_{\mu\nu\rho\lambda}\overline{q}\gamma_5 \gamma_\rho D_\lambda q$
vanishes due to equation of motion for massless quarks. \\
\begin{figure}[h]
\epsfxsize=4.5cm
\epsfbox{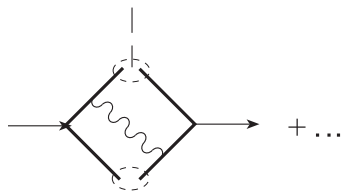}
\caption{\small{The diagrams, corresponding to the vacuum expectation value
$\langle\overline{q}\sigma_{\rho\lambda}q\rangle_F
\langle\overline{q}q\rangle$. Dots stand for 
permutations.}}
\end{figure}
\\
\indent Among the diagrams, corresponding
to the vacuum expectation values of dimension 6 operators 
(there are five such vacuum expectation values), 
the dominating contribution 
appears from the no loop diagrams with hard gluon exchange. 
In our case such diagrams contain the vacuum expectation value
$\langle\overline{q}\sigma_{\rho\lambda}q\rangle_F
\langle\overline{q}q\rangle$ (fig.3). 
Their contribution to the polarization operator is equal to
%10
$${2\over 9}{g^2\qq\chi\over{p^6}}\,.\eqno(10)$$
In this expression the quark condensate magnetic 
susceptibility $\chi$ is negative. \\
\indent The contribution of the vacuum expectation value
$\langle G^n_{\sigma\tau}G^n_{\sigma\tau}\rangle F_{\mu\nu}$ to 
$\Pi(p^2)$ is
determined by the one-loop diagrams (fig.4) and, therefore, has
numerically small factor. Since these diagrams have infrared divergence, in
order to evaluate them we introduce the cut-off over transversal momenta
$\lambda$ and obtain: 
%11
$$-{1\over{36}}\langle{\alpha_s\over\pi}G^2\rangle
\Big({1\over{2\lambda^4p^2}}-{1\over{6\lambda^2p^4}}+{3\over p^6}\Big)\,.
\eqno(11)$$
As we shall see, (11) 
is really small in comparison with (10) for any reasonable $\lambda$. \\
\begin{figure}[h]
\epsfxsize=8.5cm
\epsfbox{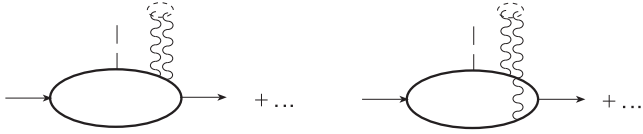}
\caption{\small{The diagrams, corresponding to the vacuum expectation value
$\langle G^n_{\sigma\tau}G^n_{\sigma\tau}\rangle F_{\mu\nu}$.}}
\end{figure}

\indent The infrared divergence is probably canceled 
by the contribution of three other
vacuum expectation values of dimension 6 operators:
%12
$$g\langle\overline{q}\Big((G^n_{\mu\lambda}D_\nu-
\overleftarrow{D}_\nu G^n_{\mu\lambda})-(G^n_{\nu\lambda}D_\mu-
\overleftarrow{D}_\mu G^n_{\nu\lambda})\Big)
\gamma_\lambda t^n q\rangle_F\,,$$
$$\epsilon_{\mu\nu\rho\lambda}g\langle\overline{q}(G^n_{\rho\xi}D_\lambda+
\overleftarrow{D}_\lambda G^n_{\rho\xi})
\gamma_\xi \gamma_5 t^n q\rangle_F\,,~~~~
d^{ikl}g^3(G^i_{\mu\lambda}G^k_{\lambda\rho}G^l_{\rho\nu}-
G^i_{\nu\lambda}G^k_{\lambda\rho}G^l_{\rho\mu})\rangle_F\,.\eqno(12)$$
Here $D_\mu$ is the covariant derivative, $d^{ikl}$ are SU(3) structure 
constants, subscript $F$ denotes the presence of the external field.
Usually vacuum expectation values of such operators can be calculated 
by constructing the corresponding sum rules.
%An example of the such approach can be found in \cite{og-sams}
%for dimension 4 operators and symmetric tensor field.
But in our case this way is inapplicable 
because of their high
dimension. However, the vacuum expectation values (12) are suppressed by 
the factor $N_c^{-1}$ ($N_c$ is the color number) as compared with
$\langle\overline{q}\sigma_{\rho\lambda}q\rangle_F
\langle\overline{q}q\rangle$, and we disregard it. \\
\indent Collecting the expressions (9), (10), (11), we find the operator
product expansion part of the sum rule:
%13 
$$\displaylines{\Pi(p^2)=-{3\over{16\pi^2}}(1+{10\over 9}{\alpha_s\over\pi})
\int\limits_0^\infty{ds\over{(s-p^2)^2}}+\hfill}$$
$$\displaylines{\hfill+{2\over 9}{g^2\qq\chi\over{p^6}}
-{1\over{36}}\langle{\alpha_s\over\pi}G^2\rangle
\Big({1\over{2\lambda^4p^2}}-{1\over{6\lambda^2p^4}}+{3\over p^6}\Big)\,.
~~~~(13)}$$
\indent Now we have both parts of required sum rule.
%ttttttttt
\section{Results and discussion}
$\,$\indent After Borel transformation 
$$\hat{B}(M^2)=\lim_{P^2,n \to \infty \atop{P^2/n=M^2}}
{(P^2)^{n+1}\over{n!}}\bigg(-{d\over{dP^2}}\bigg)^n\,,~~~~P^2=-p^2>0$$
we equate the phenomenological (8) and operator product expansion (13) parts
of the sum rule and obtain:
%14
$$\displaylines{\mu-{1\over 2}+CM^2={3g_\rho^2M^2\over{8\pi^2m_\rho^2}}
\Big(1+{10\over 9}{\alpha_s(M^2)\over\pi}\Big)
\Big(1-e^{-s_\rho/M^2}\Big)e^{m_\rho^2/M^2}-\hfill}$$
$$\displaylines{\hfill-{g_\rho^2\over{m_\rho^2}}e^{m_\rho^2/M^2}
\Bigg(-{2g^2\qq\chi\over{9M^2}}+
{1\over 36}\langle{\alpha_s\over\pi}G^2\rangle\Big({M^2\over{\lambda^4}}+
{1\over{3\lambda^2}}+{3\over{M^2}}\Big)\Bigg)\,.~~~~(14)}$$
Here $C$ appears due to nondiagonal transitions. \\
\indent We use the following values of parameters: \\
%$\mu=1\,GeV-$ the operator expansion normalization point, \\
$m_\rho=0.77\,GeV-$the $\rho$-meson mass, \\
$g_\rho^2/(4\pi)=1.27-$the $\rho$-$\gamma$ coupling constant, \\
$s_\rho=1.5\,GeV^2-$the continuum threshold for $\rho$-meson,\\
$\langle(\alpha_s/\pi)G^2\rangle=0.009\pm 0.007\,GeV^4-$the gluon condensate  
\cite{iof0207191}, \\
$g^2\qq=(0.28\pm 0.09)\times 10^{-2}\,GeV^6-$the quark condensate 
\cite{iof0207191}, \\
$\chi=-(5.7\pm 0.6)\,GeV^{-2}-$the quark condensate magnetic 
susceptibility \cite{bel-kog},\\
$\Lambda_{QCD}=0.2\,GeV$.\\
\indent In (14) $\alpha_s(M^2)=4\pi/(9\ln(M^2/\Lambda_{QCD}^2))$.
For the cut-off over transversal momenta $\lambda$ we take the value 
$\lambda^2=0.8\,GeV^2$. \\
\indent As was mentioned in \cite{sams-mrho}, in order to obtain 
the $\rho$-meson magnetic moment
in the parton model approximation  one has to substitute the relation 
for $g_\rho$ \cite{svz} 
$${g_\rho^2M^2\over{4\pi^2m_\rho^2}}\Big(1-e^{-s_\rho/M^2}\Big)
e^{m_\rho^2/M^2}=1$$
into the contribution of the bare loop and continuum (see (14)): 
$$\mu_{part}-{1\over 2}={3g_\rho^2M^2\over{8\pi^2m_\rho^2}}
\Big(1-e^{-s_\rho/M^2}\Big)e^{m_\rho^2/M^2}.$$
It was obtained in this way:
$$\mu_{part}=2\,.$$
This result agrees with the prediction of the vector dominance hypothesis. \\
\indent Now let us analyze the whole equation (14). 
In order to find the value of the magnetic moment, 
we approximate the right-hand 
side of (14) $R(M^2)$ (see fig.5) by a straight line in the  interval
$1.0\,GeV^2 \le M^2\le 1.3\,GeV^2$ and find it ordinate
at zero Borel mass. \\ 
% Another way consists in applying
%of the operator $1-M^2(d/dM^2)$ to equations (26), (27). \\
\indent Thus we obtain:
$$\mu=1.8\,.$$
\begin{figure}[h]
\epsfxsize=11.0cm
\epsfbox{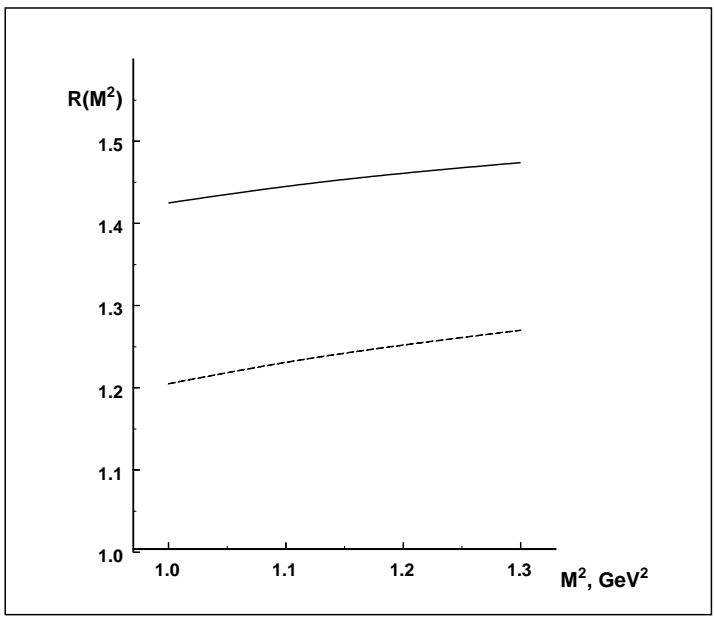}
\caption{\small{The right-hand side of equation (14) $R(M^2)$ 
as the function of $M^2$ (solid line). The same without 
$\alpha_s$-corrections (dashed line).}}
\end{figure}
%Within the chosen interval of $M^2$ the graph of the right-hand side of (14)
%is close to the straight line (see fig.1).\\
\indent The operators of dimension 6 give less than 20\% to this value. \\
\indent The contribution of the gluon condensate does not exceed $30\%$ 
of the total contribution of dimension 6 operators,
the contribution of the terms, which contain $\lambda^2$, is 
$\la 15\%$ of the same quantity. 
That is why variation of 
$\lambda^2$ within the interval $0.6\,GeV^2 \le \lambda^2\le 1.0\,GeV^2$ 
does not change the value of the magnetic moment. \\
\indent The variations of the values of the quark and gluon condensates
within the given limits change the value of magnetic moment by 
$\la 10\%$ each. \\ 
\indent The uncertainty in the value of the quark condensate magnetic 
susceptibility results in the error about few percent 
in the value of the magnetic
moment. Variation of the continuum threshold for the $\rho$-meson $s_\rho$ 
in the reasonable limits gives the same effect. \\ 
Taking into account all uncertainties we obtain: 
%15
$$\mu=1.8\pm 0.3\,.\eqno(15)$$
This is our final result. \\
\indent In \cite{iof-3vertex} it was shown that the approximation procedure
is correct (nonlinear terms can be safely neglected), when $CM^2/\mu\ll 1$.
In our case $CM^2/\mu\approx 0.2$.\\
\indent Thus we see that in parton model (bare loop) approximation 
the  $\rho$-meson magnetic moment $\mu_{part}=2$. All accounted 
operator product expansion corrections are negative, i.e. nonperturbative
interactions decrease this quantity. We obtain that 
the value of the magnetic moment (disregarding $\alpha_s$-corrections) 
is equal to $1.5$ \cite{sams-mrho}.
On the other hand, the $\alpha_s$-correction
is quite appreciable ($\sim 0.3$) in magnitude and positive (see fig.5). 
After taking into account both perturbative and nonperturbative terms, 
we obtain the value (15). It is consistent with the vector 
dominance model prediction again. \\
\indent The value of the $\rho$-meson magnetic moment was calculated in a
number of papers within the Dyson-Schwinger equation 
based models. In \cite{hawes} the value $\mu=2.69$ was found. In \cite{hecht} 
several results are compared, and the values of $\mu$ 
lie between 2.5 and 3.0. Relativistic quantum mechanics model
gives \cite{rel-qm} $\mu=2.23\pm 0.13$. \\
\indent Recently the value of the magnetic
moment was found in the light cone QCD sum rule technique \cite{aliev-kanik}.
In this technique the expansion over the twist of the operators is performed,
and all nonperturbative effects are encoded in the wave functions. In 
\cite{aliev-kanik} the asymptotic form $\psi(u)=1$ of the twist 2 photon wave
function $\psi(u)$ at the point $u=1/2$ was used, and it was obtained 
(disregarding $\alpha_s$-corrections): $\mu=2.2\pm 0.2$. This value and (15)
do not contradict one another within the errors.
\\ \\
\indent The author is thankful to B.L.\,Ioffe for the statement of the 
problem and valuable discussions 
and to A.G.\,Oganesian for helpful discussions. \\
\indent The work is supported in part by grants INTAS 2000
Project 587 and RFFI 03-02-16209.


\begin{thebibliography}{99}
\bibitem{kroll} N.\,Kroll, T.\,Lee and B.\,Zumino, Phys.Rev. 157 (1967)
 1376.
\bibitem{iof-kh-lip} B.\,Ioffe, V.\,Khoze and L.\,Lipatov,
Hard Processes, North Holland, 1984, ch.5. 
\bibitem{iof-sm-meson} B.\,Ioffe and A.\,Smilga, Nucl.Phys. B216 (1983) 373.
\bibitem{sams-mrho} A.\,Samsonov, hep-ph/0208165, submited to Yad.Fiz.
\bibitem{iof-sm-magmom} B.\,Ioffe and A.\,Smilga, Nucl.Phys. B232 (1984) 109.
\bibitem{bal-yung} I.\,Balitsky and A.\,Yung, Phys.Lett. B129 (1983) 328.
\bibitem{hawes} F.\,Hawes and M.\,Pichowsky, Phys.Rev. C59 (1999) 1743.
\bibitem{hecht} M.\,Hecht and B.H.J.\,McKellar, Phys.Rev. C57 (1998) 2638.
\bibitem{rel-qm} J.P.B.C de Melo and T.\,Frederico, Phys.Rev. C55 
(1997) 2043.
\bibitem{aliev-kanik} T.\,Aliev, I.\,Kanik, M.\,Savci, hep-ph/0303068.
%\bibitem{og-sams} A.\,Oganesian and A.\,Samsonov, JHEP 0109 (2001) 002.
\bibitem{chetyrkin} K.\,Chetyrkin and F.\,Tkachov, 
Nucl.Phys. B192 (1981) 159.
\bibitem{iof0207191} B.\,Ioffe, Phys.Atom.Nucl. 66 (2003) 30,
 Yad.Fiz. 66 (2003) 32.
\bibitem{bel-kog} V.\,Belyaev and Ya.\,Kogan, Yad.Fiz. 40 (1984) 1035
(Sov.J.Nucl.Phys. 40 (1984) 659).
%\bibitem{iof-zyab} B.\,Ioffe and K.\,Zyablyuk, Nucl.Phys. A687 (2001) 437.
\bibitem{svz} M.\,Shifman, A.\,Vainshtein and V.\,Zakharov, 
Nucl.Phys. B147 (1979) 385, 448.
\bibitem{iof-3vertex} B.\,Ioffe, Yad.Fiz. 58 (1995) 1492.
\end{thebibliography}
\end{document}